\documentclass{PoS}
\usepackage{epsf}  
\title{Rapidity Variation of Thermal Parameters at SPS and RHIC.
\thanks{
Work supported  by the Scientific and
Technological Co-operation Programme between 
Italy and South Africa, project number 16.  }}
\ShortTitle{Rapidity Variation.}
\author{F. Becattini\\
 Universit\`a di Firenze and INFN Sezione di Firenze \\
Largo E. Fermi 2, I-50125, Florence, Italy.\\
        E-mail: \email{becattini@fi.infn.it}}
\author{\speaker{ J. Cleymans} and J.~Str\"umpfer
\\
UCT-CERN Research Centre and Department  of  Physics, \\
Rondebosch 7701, Cape Town, South Africa\\
E-mail: \email{Jean.Cleymans@uct.ac.za}}
\abstract 
{
The rapidity dependence of the 
chemical freeze-out thermal parameters $T$ and $\mu_B$ are 
determined at the highest RHIC and  SPS energies.
 These show a systematic behavior towards an increase in $\mu_B$
away from mid-rapidity and a corresponding decrease in the 
temperature $T$.
}

\FullConference{Critical Point and Onset of Deconfinement
          4th International Workshop\\
		 July 9-13 2007\\
		 GSI Darmstadt,Germany}

\begin{document}
Data from the BRAHMS collaboration~\cite{BRAHMS-shanghai} show that 
antiparticle to particle ratios have a maximum at mid-rapidity and
slowly decrease as the rapidity becomes larger.
It was emphasized very eloquently  by R\"ohrich~\cite{roehrich} that 
the particle ratios at 
large rapidities  are consistent with those 
measured at the SPS energies. 
This opens the possibility to compare
measurements for e.g. the $K^+/\pi^+$ ratio at high rapidities and check them
with the corresponding values measured in the energy scan at the SPS, thus 
complementing information about the rapid variation of this ratio 
as a function of
beam energy. 
The sharp variation in this ratio with beam energy
remains a mystery for most models and might indicate interesting dynamics 
possibly even linked to the appearance of the critical point at intermediate values
of the baryon chemical potential in 
lattice quantum chromodynamics~\cite{karsch,fodor}.

A change in particle - anti-particle ratios with longitudinal momentum
has been observed in many collisions, it was first discussed two decades ago in an
analysis of $p-p$ and  $p-N$ data~\cite{greiner}.  

\begin{figure}[thb]
\centerline{
\includegraphics[width=75mm,height=62mm,clip]{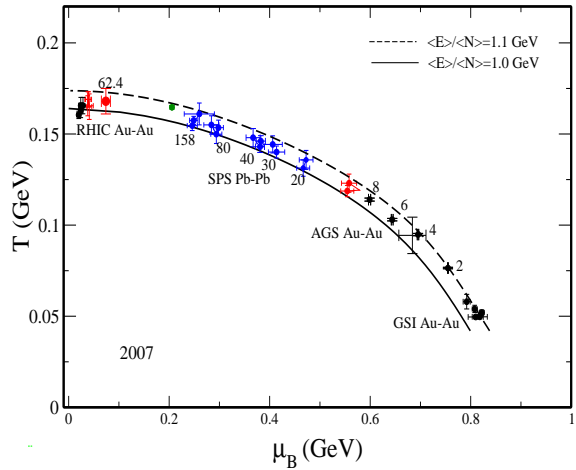}}
\caption{ Temperature vs. $\mu_B$ as determined from heavy ion 
collisions at different beam energies. The lower AGS points are 
based on a preliminary analysis of $4\pi$ data.
The  RHIC point at 62.4 GeV was obtained by J. Takahashi~\cite{takahashi}.
See~\cite{prc06,manninen} for more details.}
 \label{eovern}
\end{figure}

The analysis of particle 
multiplicities in heavy ion collisions has shown overwhelming
evidence for chemical equilibrium in the final state except for
particles carrying strangeness which are  suppressed at lower beam energies; however, their 
relative yields fulfill statistical equilibrium.
A summary  as of 2007, combining the results from many
different groups,  is shown in Fig. \ref{eovern}.
Except for particle multiplicities at RHIC energies, all data
in Fig.~\ref{eovern} use integrated particle yields, the
very systematic change of thermal parameters over the full range 
of beam energies is one of the most impressive features 
of relativistic ion collisions to date.
It is now possible to use the thermal model to make reliable predictions for
particle multiplicities at LHC energies~\cite{kraus} and to determine
which beam energy will lead to the highest baryon density at 
freeze-out~\cite{randrup}. 
With chemical equilibrium thus firmly 
established,  we
focus on other properties, in particular,
since the rapidity distributions of identified particles is  now  becoming 
available also at RHIC energies~\cite{BRAHMS-shanghai},  it is now 
possible to determine the rapidity dependence of thermal parameters.
A first analysis was done by Stiles and Murray~\cite{stiles} for the 
data obtained by the
BRAHMS collaboration
at 200 GeV~\cite{BRAHMS-shanghai}. 
This shows a clear dependence of 
the baryon chemical potential on rapidity  due to the 
changing $\bar{p}/p$ ratio.
A thorough analysis of the rapidity dependence was recently done 
in Ref.~\cite{broniowski1,broniowski2} using a model based on a 
single freeze-out temperature. 
A first report of our results was  presented elsewhere~\cite{beca_jc}.

The general procedure is as follows: the rapidity axis is populated
with fireballs following a gaussian distribution function given by
$\rho(y_{FB})$ where $y_{FB}$ is the rapidity of the fireball.
\begin{equation}
\rho(y_{FB}) = \frac{1}{\sqrt{2}\pi\sigma} 
\exp\left( -\frac{y^2_{FB}}{2\sigma^2}\right) .
\label{eqrho}
\end{equation}

 Particles 
will appear when the fireball freezes out and  will 
follow a thermal distribution  centered around the position
of the fireball 
The momentum distribution of hadron $i$ is then calculated from 
the distribution of fireballs as given by Eq.~[\ref{eqrho}] along 
the rapidity axis as follows 
\begin{equation}
E_i\frac{d^3N_i}{d^3p} = \int_{-\infty}^{\infty} 
\rho\left(y_{FB}\right)E_i\frac{d^3N_1^i}{d^3p}(y-y_{FB})~dy_{FB}
\label{eqdist}
\end{equation}
where
$E_i\frac{d^3N_1^i}{d^3p}$ is the distribution of hadrons 
from a single fireball.
The temperature $T$ and the baryon chemical potential $\mu_B$ will 
depend on the 
rapidity of the fireball and are not assumed to be constant.

We have included  resonance decays in the final 
distribution and assumed they decay isotropically.

\begin{figure}[thb]
\centerline{
\includegraphics[width=75mm,height=62mm,clip]{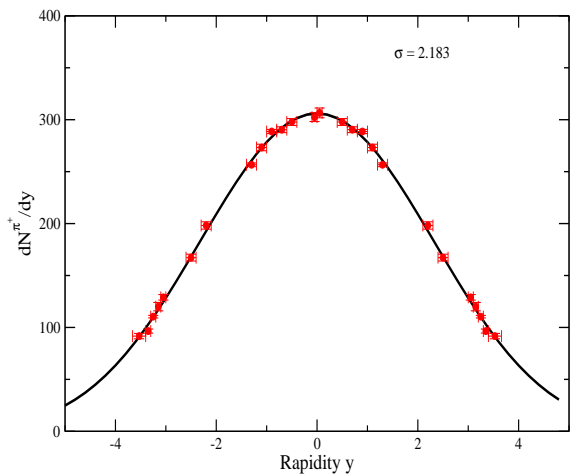}}
\caption{Fit to the pion distribution as measured by the BRAHMS
collaboration.}
 \label{fig:pions}
\end{figure}

An important parameter is the width of the distribution.
For the RHIC data at 200 GeV this  was determined from the 
  $\pi^+$'s as these are 
very sensitive to the value of $\sigma$ and  less to variations in
the baryon chemical potential. 
The width of the distribution 
$\sigma = 2.183$  is compatible with the values quoted by
the  BRAHMS collaboration~\cite{BRAHMS-shanghai},  e.g.
$\sigma_{\pi^+} = 2.25\pm 0.02$ 
and 
$\sigma_{\pi^-} = 2.29\pm 0.02$. This is shown in Fig.~\ref{fig:pions}.
The hadrons described by Eq.~[\ref{eqdist}] are mainly hadronic resonances.
Only a fraction of these are stable under strong interactions and 
most of them decay
into stable hadrons at chemical freeze-out, hence the need to implement 
multi-particle decays.

\begin{figure}[thb]
\centerline{
\includegraphics[width=75mm,height=62mm,clip]{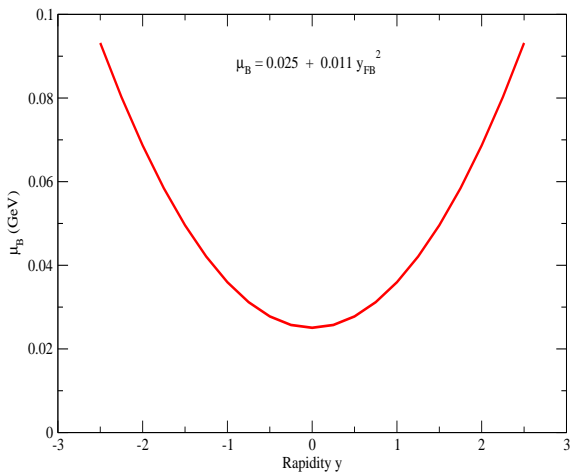}}
\caption{Values of the baryon chemical potential as a 
function of rapidity at the highest RHIC energy.}
 \label{fig:muy200}
\end{figure}

We assume that the temperature  $T$ and the chemical potential
$\mu_B$ are always related 
via the freeze-out curve as given in Fig.~[\ref{eovern}]; if the temperature 
varies along the rapidity axis, then also the
 chemical potential will vary. Thus a decrease in the temperature
of the fireball will be accompanied by an increase in the 
baryon chemical potential. 
In other words, we assume a universality of the chemical freeze-out condition.
This relationship between 
temperature and baryon chemical potential 
is  very reasonable since all particle abundances measured so far 
follow it.
Once the width of the distribution of fireballs has been fixed, we can 
go on with the
dependence of the baryon chemical potential on the rapidity of the fireball
(units are in GeV) at RHIC:
\begin{equation}
\mu_B = 0.025 + 0.011~y_{FB}^2
\end{equation}
This is shown graphically in Fig.~[\ref{fig:muy200}].

\begin{figure}[thb]
\centerline{
\includegraphics[width=75mm,height=62mm,clip]{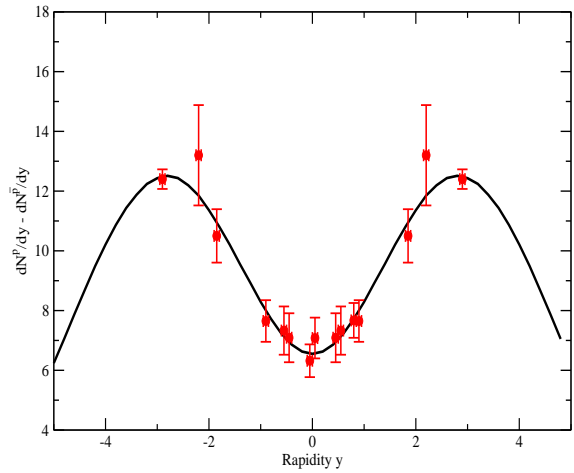}}
\caption{ The  $p - \bar{p}$ distribution as a function of rapidity. 
}
 \label{fig:netp_cpod_200}
\end{figure}

A comparison of the resulting
 net proton  distribution $p - \bar{p}$   
with  RHIC data at 200 GeV is shown in Fig.~[\ref{fig:netp_cpod_200}].

\begin{figure}[thb]
\centerline{
\includegraphics[width=75mm,height=62mm,clip]{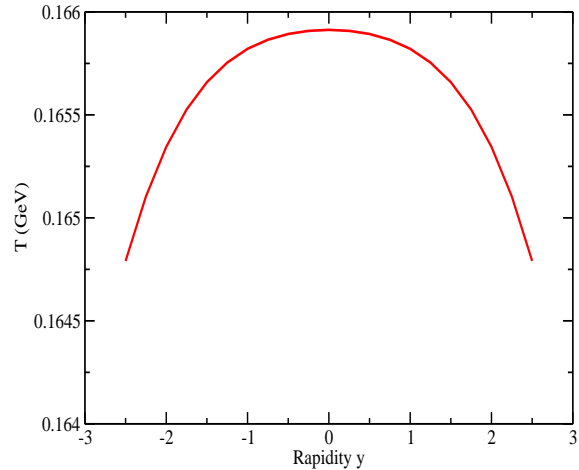}}
\caption{Values of the chemical freeze-out temperature  
function of rapidity at the highest RHIC energy.}
 \label{fig:ty200}
\end{figure}
The variation of the temperature along the rapidity axis is shown in 
Fig.~[\ref{fig:ty200}]. The temperature is maximal at mid-rapidity
and gradually decreases towards higher (absolute) values of the 
rapidity. Note that the temperature varies less than 2 MeV over the 
rapidity interval.

\begin{figure}[thb]
\centerline{
\includegraphics[width=75mm,height=62mm,clip]{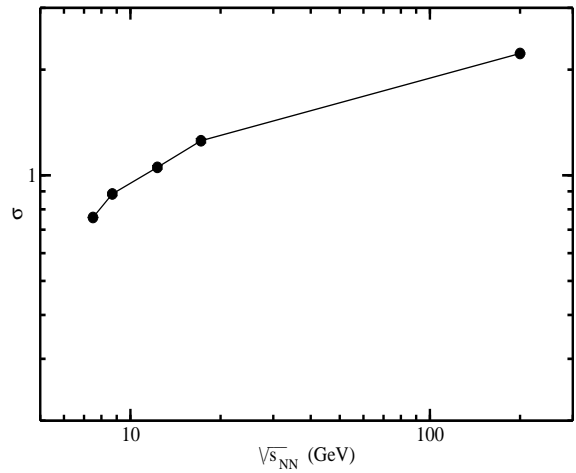}}
\caption{Width of the Gaussian distribution, $\sigma$, used to describe the
distribution of fireballs for various beam energies.
}
 \label{fig:sigma}
\end{figure}

\begin{figure}[thb]
\centerline{
\includegraphics[width=75mm,height=62mm,clip]{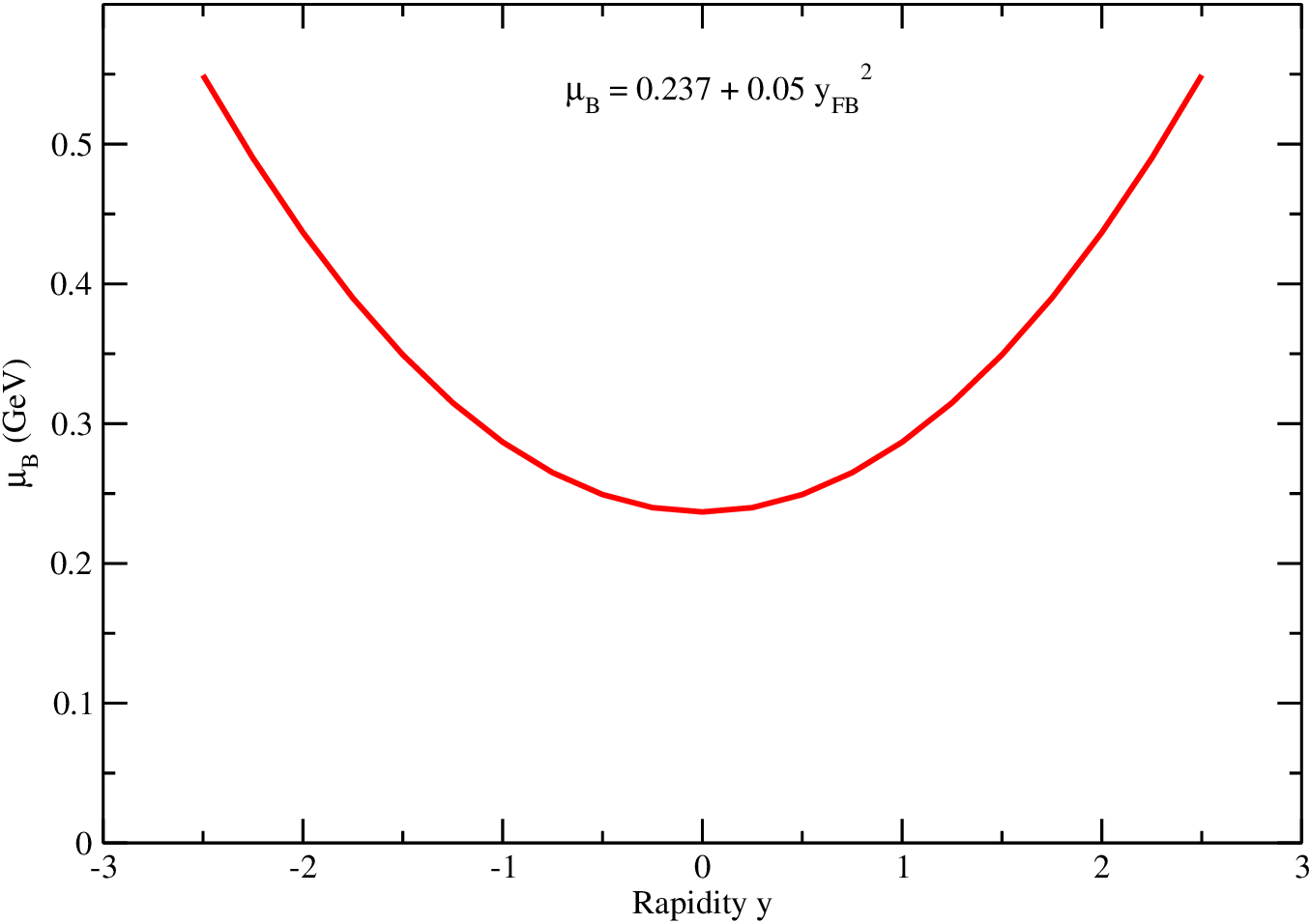}}
\caption{Values of the baryon chemical potential as a function of 
rapidity at the highest SPS  energy.}
 \label{fig:muy17}
\end{figure}

\begin{figure}[thb]
\centerline{
\includegraphics[width=75mm,height=62mm,clip]{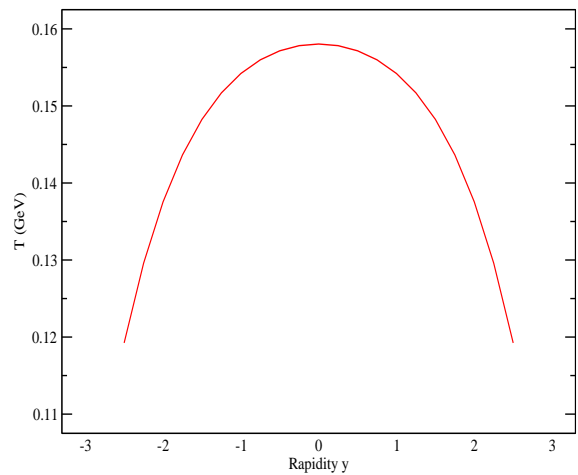}}
\caption{Values of the chemical freeze-out temperature  
as a function of rapidity at the highest SPS  energy.}
 \label{fig:ty17}
\end{figure}

The situation at the highest SPS energy is more difficult 
to describe because the changes of the 
 thermal variables with rapidity are much larger. First of all we need the 
distribution of fireballs at SPS energies. This is determined by the 
variable $\sigma$ appearing in Eq.~(\ref{eqrho}) which we fixed using the 
distribution of pions. The results obtained at various beam
energies are  shown in Fig.~[\ref{fig:sigma}].
\begin{figure}[thb]
\centerline{
\includegraphics[width=75mm,height=62mm,clip]{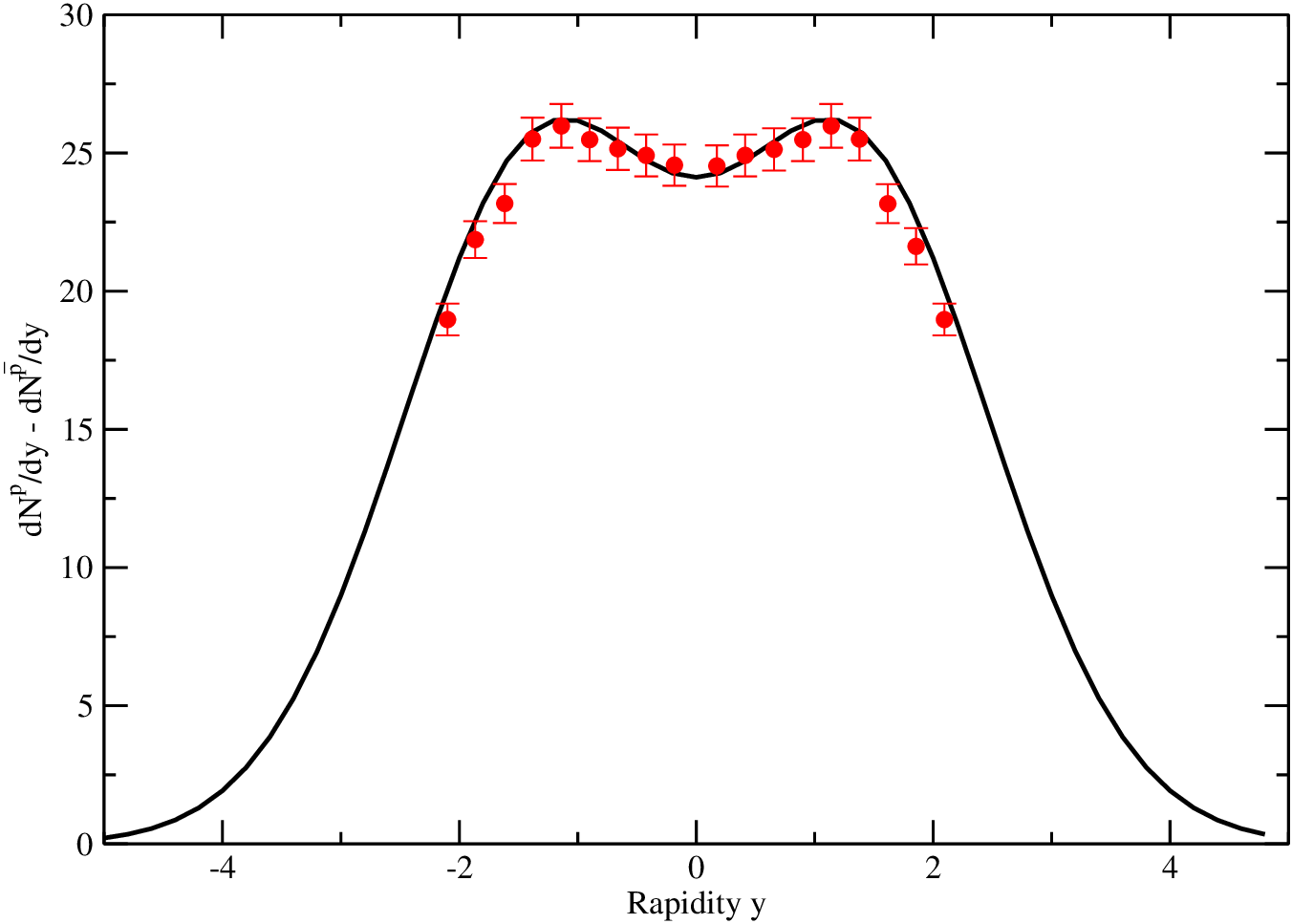}}
\caption{ The $p - \bar{p}$ distribution as a function of rapidity
at SPS. }
 \label{fig:netp_cpod_17}
\end{figure}
The baryon chemical potential can approximately be described by:
\begin{equation}
\mu_B = 0.237 + 0.011~y_{FB}^2  .
\end{equation}
This is shown graphically in Fig.~[\ref{fig:muy17}].
The corresponding change in the temperature as a function of rapidity
is shown in Fig.~[\ref{fig:ty17}]. Whereas there is almost no change in the temperature 
in the rapidity interval under consideration (the change is less than 2 MeV at RHIC), 
at the highest SPS energy the change in temperature is much larger and ranges from about 160 MeV 
down to about 120 MeV. 
The  comparison between the resulting
 net proton  distribution $p - \bar{p}$   
with  SPS data at 17.2 GeV is shown in Fig.~[\ref{fig:netp_cpod_17}].

In summary, particle yields measured in heavy ion collisions show an overwhelming evidence for
chemical equilibrium at all beam energies. The rapidity dependence of the 
thermal parameters $T$ and $\mu_B$ can now be determined over a wide range of
rapidities and show a systematic behavior towards an increase in $\mu_B$
as the rapidity is increased.
%
 
%
\end{document}